\documentclass[amssymb,prb,twocolumn,superscriptaddress,floats,showpacs]{revtex4-2}
\usepackage[T1]{fontenc}
\usepackage{bm}
\usepackage{graphicx}
\usepackage{amssymb}
\usepackage{amsfonts}
\usepackage{amsmath} 
\usepackage{hyperref} 
\usepackage{textcomp}
\usepackage{xcolor}
\usepackage{xspace}
\hypersetup{colorlinks,citecolor=blue, filecolor=blue ,linkcolor=blue , urlcolor=blue, pdftex}
\usepackage{color}
\begin{document}

\newcommand{\MoSe}{$\text{MoSe}_{2}$\xspace}
\newcommand{\ReS}{$\text{ReS}_{2}$\xspace}

\title{Ultralow-Tensile Strain Enables Exciton Funneling and Energy Transfer to Boost \MoSe Photoluminescence Quantum Yield}

% {author & affiliation}

\def \FUW{Faculty of Physics, University of Warsaw, 02-093 Warsaw, Poland}

\def \CNRS{Université Paris-Saclay, ONERA, CNRS, Laboratoire d’étude des microstructures (LEM), F-92322 Châtillon, France}

\def \deb{School of Basic Sciences, Indian Institute of Technology Bhubaneswar, Khordha, Odisha, India}

\def \Watanabe{Research Center for Electronic and Optical Materials, National Institute for Materials Science, 1-1 Namiki, Tsukuba 305-0044, Japan}

\def \Taniguchi{Research Center for Materials Nanoarchitectonics, National Institute for Materials Science,  1-1 Namiki, Tsukuba 305-0044, Japan}

\author{Gayatri}
\email{gayatri@fuw.edu.pl}
\affiliation{\FUW}
\author{Mehdi Arfaoui}
\affiliation{SPEC, CEA, CNRS, Université Paris-Saclay, CEA Saclay, F-91191 Gif-sur-Yvette, France}
\affiliation{\CNRS}
\author{Debashish Das}
\affiliation{\deb}
\author{Mateusz Raczyński}
\affiliation{\FUW}
\author{Marta Bilska }
\affiliation{\FUW}
\author{Piotr~Tatarczak}
\affiliation{\FUW}
\author{Aleksandra~Krystyna Dąbrowska }
\affiliation{\FUW}
\author{Tomasz Kazimierczuk }
\affiliation{\FUW}
\author{Takashi~Taniguchi}
\affiliation{\Taniguchi}
\author{Kenji~Watanabe}
\affiliation{\Watanabe}
\author{Piotr Kossacki}
\affiliation{\FUW}
\author{Andrzej Wysmołek}
\affiliation{\FUW}
\author{Saroj~Kumar~Nayak}
\affiliation{\deb}
\author{Adam Babiński}
\affiliation{\FUW}
\author{Johannes Binder}
\affiliation{\FUW}
\author{Maciej R. Molas}
\affiliation{\FUW}
\author{Arka Karmakar}
\email{arka.karmakar@fuw.edu.pl}
\affiliation{\FUW}

\begin{abstract} 
Strain engineering is a powerful route for controlling the exciton dynamics in van der Waals (vdW) heterostructures (HSs). 
The interlayer energy transfer (ET) process is another key factor in controlling the photocarrier relaxation pathways in vdW HSs. 
In this work, we combine these two processes to achieve an $\sim$8-fold enhancement to the relative photoluminescence (PL) quantum yield (QY) in a HS formed from monolayers of \ReS and \MoSe, separated by a thin hBN interlayer, placed onto an hBN bubble. 
We achieve this enhancement by applying only $\sim$0.1\% biaxial tensile strain, which results in efficient exciton funneling and an increased transition dipole moment. 
Our experimental data are supported by first-principles density-functional theory and coherent transfer-matrix method calculations, ruling out optical interference as the dominant origin of the enhancement. 
This work provides an innovative route for enhancing the PL QY of vdW materials via interplay between the tensile strain and the ET process.  
\end{abstract}


\maketitle

%%%%%%%%%%%%%%%%%%%%%%% {INTRO} %%%%%%%%%%%%%%%%%%%%%%%%%%%%%%%%%
\section{Introduction \label{sec:Intro}}
Mechanical strain has emerged as an effective tool for tailoring the electronic, optical, and excitonic properties of two-dimensional (2D) van der Waals (vdW) semiconductors~\cite{Manzeli2017, Qi2023StrainEngineering,Conley2013}.
In monolayer (ML) transition metal dichalcogenides (TMDCs), local strain is commonly introduced using bubbles, wrinkles, nanopillars, or patterned substrates~\cite{Branny2017, Datta2020, StrainBubbles2019ACSPhotonics,Xiong2024MoS2BubblesHER, Ai2025}. 
In particular, strain generated by bubbles creates localized potential landscapes that promote exciton funneling and modify light emission --- providing an effective route toward tunable nanoscale optoelectronic and quantum photonic devices~\cite{Feng2012,Kumar2015, Mueller2018}.
When different TMDC MLs are vertically stacked to form heterostructures (HSs), strain also influences the interlayer coupling~\cite{Geim2013, Rivera2015,Novoselov2016}. 

Interlayer energy transfer (ET) process, particularly non-radiative Förster resonance ET provides an efficient  pathway to control the exciton dynamics in vdW HSs without relying on the interlayer charge transfer (CT) process~\cite{Kozawa2016FastET,Jin2018UltrafastDynamics, Karmakar2023ExcitationDependent, Karmakar2024TwistedMoSe2, Gayatri2026FastInterlayerET}. 
In type-II TMDC HSs, ET competes with CT and strongly depends on the donor-acceptor spectral overlap, and interlayer separation~\cite{Kozawa2016FastET, Liu2021EnergyChargeTransfer, Karmakar2022I, Aftab2023EnergyChargeTransfer, Gayatri2026FastInterlayerET}. 
A recent experimental study  has shown the effect of compressive strain on the ET process by placing a TMDC HS inside a diamond anvil cell ~\cite{Kim2022StrainCTET}. 
However, the influence of the small tensile strain that inevitably arises from bubble formation during the HS fabrication remains experimentally unexplored. Thus, understanding how weak local tensile strain modifies the interlayer exciton dynamics is timely and practically relevant.
%However, despite these previous studies~\cite{Kozawa2016FastET, Liu2021EnergyChargeTransfer, Karmakar2022I, Kim2022StrainCTET, Aftab2023EnergyChargeTransfer, Gayatri2026FastInterlayerET}, the influence of small tensile strain induced by bubble formation on the ET process has yet to be experimentally established. Especially when bubble formation is unavoidable during the HS fabrication steps, a comprehensive understanding of how weak local strain modifies the interlayer exciton dynamics becomes timely and necessary. %However, despite all these previous studies~\cite{Kozawa2016FastET, Liu2021EnergyChargeTransfer, Karmakar2022I, Kim2022StrainCTET, Aftab2023EnergyChargeTransfer, Gayatri2026FastInterlayerET}, an experimental understanding about the effect of small tensile strain on the ET process by bubble formation, is yet to be reported.
In this work, we demonstrate that only $\sim$0.1\% biaxial tensile strain produces an $\sim$8-fold enhancement of the relative \MoSe photoluminescence (PL) quantum yield (QY) in an all-ML \ReS/hBN/\MoSe HS, by placing the entire stack onto a hBN bubble/sapphire substrate. The thin ($\sim$12 nm) hBN spacer suppresses the CT process, while enabling the long-range dipole–dipole ET~\cite{Karmakar2022I}. 
%In this work, we demonstrate an efficient ET process in a type-II HS, formed from ML of rhenium disulfide (ReS$_2$) and ML of molybdenum diselenide (MoSe$_2$), separated by a thin ($\sim$12~nm) layer of hexagonal boron nitride (hBN) spacer. The entire stack was placed on an hBN bubble to induce local biaxial tensile strain. 
The strain simultaneously drives the exciton funneling and increases the transition dipole moments (TDMs) of both MLs --- boosting the ET efficiency. First-principles calculations and transfer-matrix modeling (TMM) support the experimental findings and rule out the possibility of optical interference as the major contributor of the observed enhancement.
%promoting efficient dipole-dipole interactions and exciton funneling~\cite{Cheng2020, Priyanka2024, StrainBubbles2019ACSPhotonics}. 
Importantly, the present results are obtained in a purely ML HS; increasing the optical absorption of the donor layer is therefore expected to yield even larger gains in \MoSe emission. 
Our findings also promise to further increase the PL quantum yield (QY) of any TMDC HS by applying a higher strain in a controllable way --- ideal for real vdW optoelectronic device applications.

%%%%%%%%%%%%%%%%%%%%%%% {Experimental details}
\section{Results and Discussion\label{sec:Results}}

%%%%%%%%%%%%%%%%%%%%%%% {Sample Fabrication and Strain Platform

\begin{figure*}[t]
    \centering
    \includegraphics[width=1\linewidth]{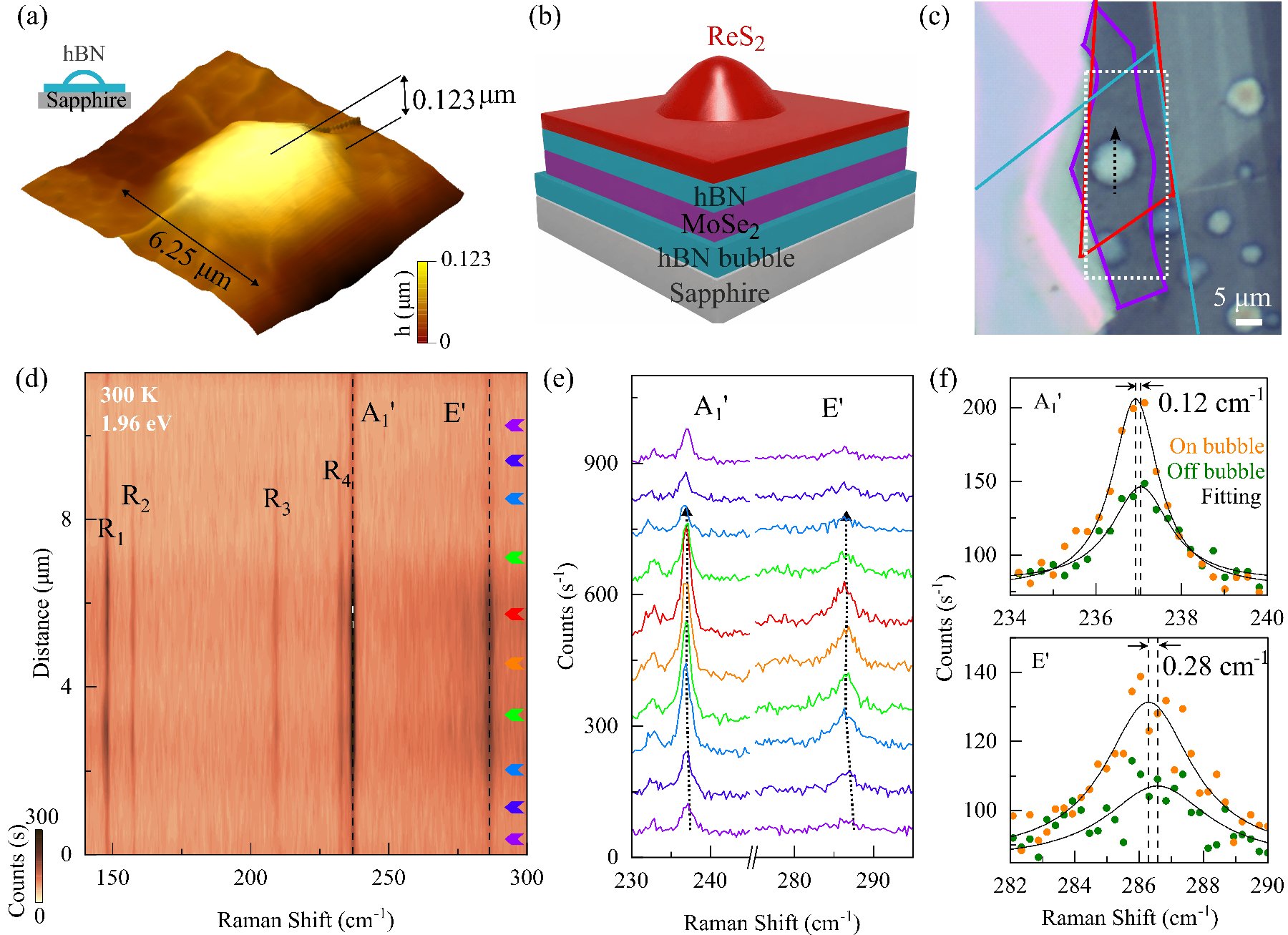}
    \caption
    {\label{fig_1}
    (a) Three-dimensional AFM image of the hBN bubble on the sapphire substrate, showing a height of $\sim$0.12~$\mu$m and base diameter of  $\sim$6.25~$\mu$m.  
    (b) Schematic illustration of the HS cross-section (from top): ML \ReS, interlayer hBN, ML \MoSe and the bottom hBN bubble/sapphire substrate. 
    (c) Optical image of the sample. Colored outlines correspond to the layers shown in (b). 
    (d) RT Raman intensity map along the black dotted arrow shown in (c) measured using 1.96 eV excitation. Overall, Raman intensity on the bubble is significantly higher as compared to the outside of the bubble. 
    (e) Selected Raman spectra from different positions along the line scan shown in (d) --- showing small shifts in \MoSe $A_1^{\prime}$ and $E^{\prime}$ modes at the bubble apex. Vertical offsets are added to the spectra for a better visual representation.
    (f) Top and bottom panels show zoomed in \MoSe Raman peaks --- then Lorentzian fits are applied to to determine the peak shifts from the on-bubble (strained) and off-bubble (unstrained) positions. The extracted redshifts from the unstrained to strained region are $|\Delta\omega_{A_1^{\prime}}| = 0.12$~cm$^{-1}$ and $|\Delta\omega_{E^{\prime}}| = 0.28$~cm$^{-1}$.
    }
\end{figure*}

The investigated HS is assembled layer-by-layer on a specially prepared hBN bubble/sapphire substrate, which provides a well-defined platform for strain engineering~\cite{Binder2023Epitaxial}. 
A detailed discussion of the hBN growth, bubble formation and HS fabrication is provided in the methods section.
Figure~\ref{fig_1}(a) shows a three-dimensional atomic force microscopy (AFM) image of the hBN bubble on the sapphire substrate after transferring ML \MoSe. 
The bubble exhibits a height of 0.12~$\mu$m and a base diameter of 6.25~$\mu$m. 
Following the bubble profile reported by Khestanova \textit{et al.}~\cite{Khestanova2016UniversalShape} and the geometrical approach for estimating strain from the bubble morphology~\cite{Hwang2022ShearStrainPL, Blundo2020Engineered}, the geometrical strain is estimated to be $\sim$0.3\% [calculations shown in Section S1 of the Supplementary Information (SI)].
This estimated value represents an upper bound, since elastic relaxation during bubble formation lowers the actual strain transferred to the MLs. 
The actual strain present in the transferred HS is therefore further verified independently by Raman and PL measurements as discussed in the following section. 
Figure~\ref{fig_1}(b) presents a schematic cross-sectional view of the sample and Fig.~\ref{fig_1}(c) shows the optical image of the sample, in which the bubbles appear as bright spots and the layer boundaries are color-coded consistently with the schematic. 
The thickness of the interlayer hBN of $\sim$12~nm (AFM height profile is shown in SI Section S2) is chosen to completely suppress the CT process~\cite{Karmakar2022I}.

%%%%%%%%%%%%%%%%%%%%%%%{Strain Characterization by Raman Spectroscopy} 
To characterize the strain distribution across the bubble, room temperature (RT) Raman line scans under 1.96~eV excitation energy are performed [Fig.~\ref{fig_1}(d)] from the flat HS region towards the apex of the bubble along the direction indicated by the black dotted arrow in Fig.~\ref{fig_1}(c). 
Four \ReS Raman-active modes ($R_1$--$R_4$) are identified at 148.03, 157.39, 209.18, and 232.85~cm$^{-1}$. 
The \MoSe $A_1^{\prime}$ and $E^{\prime}$ modes are observed at 236.92 and 286.29~cm$^{-1}$, respectively. 
Individual Raman spectra are shown in SI Section S3. 
The map reveals a clear enhancement of Raman intensity in the strained region for all modes, indicating an efficient ET process~\cite{Dandu2020}.
%%indicating the possibility an efficient ET process between the layer
For a quantitative analysis of the strain magnitude and character, we focus on the \MoSe $A_1^{\prime}$ and $E^{\prime}$ modes, due to their higher symmetry compared with the the low-symmetry \ReS modes. 
Representative \MoSe Raman modes at several positions along the Raman map are presented in Fig.~\ref{fig_1}(e). 
Both modes exhibit a systematic increase in intensity and a small but measurable redshift when moving from the flat region to the bubble apex. 
The $E^{\prime}$ mode shows a larger shift than the $A_1^{\prime}$ mode, consistent with its greater sensitivity to strain~\cite{Yagmurcukardes2018, Pak2017}.
The nature of the strain is first assessed \textit{via} the intensity ratio $I_{E^{\prime}}/ I_{A_1^{\prime}}$. 
In the unstrained region, this ratio is $0.49$ and increases to $0.57$ over the bubble --- demonstrating the presence of biaxial tensile strain~\cite{Yagmurcukardes2018}.

For a numerical determination of the \MoSe Raman shifts, Lorentzian fits are applied to spectra extracted from the on-bubble (strained) and off-bubble (unstrained) positions, as shown in Fig.~\ref{fig_1}(f).
The extracted redshifts from the unstrained to strained region are $|\Delta\omega_{A_1^{\prime}}| = 0.12$~cm$^{-1}$ and $|\Delta\omega_{E^{\prime}}| = 0.28$~cm$^{-1}$. 
These redshifts are consistent with the phonon softening due to the elongation of the Mo–Se bonds under a tensile strain, which reduces the interatomic force constants~\cite{Chang2013}.
The strain magnitude is estimated using the biaxial Raman shift coefficients reported for \MoSe in Ref.~\cite{Yagmurcukardes2018} (details are in SI Section S1). The calculated data shows a consistent picture of a weak local biaxial tensile strain between $\sim$$0.09$–$0.12\%$, matching well with earlier reports~\cite{Qian2025}. 
An independent estimation derived from the PL peak redshift taken along the arrow in Fig.~\ref{fig_1}(c) (see SI Section S4) indicates a tensile strain $\sim$$0.06\%$. 
This slight difference in the strain value is expected due to the high-sensitivity of the Raman modes for lattice distortions~\cite{Yagmurcukardes2018}. Due to the limited spatial resolution of the optical setup, the discussion is restricted to the flat (off-bubble) region and the bubble apex; intermediate sidewall regions are not analyzed.

\begin{figure}[t!]
    \centering
    \includegraphics[width=1\linewidth]{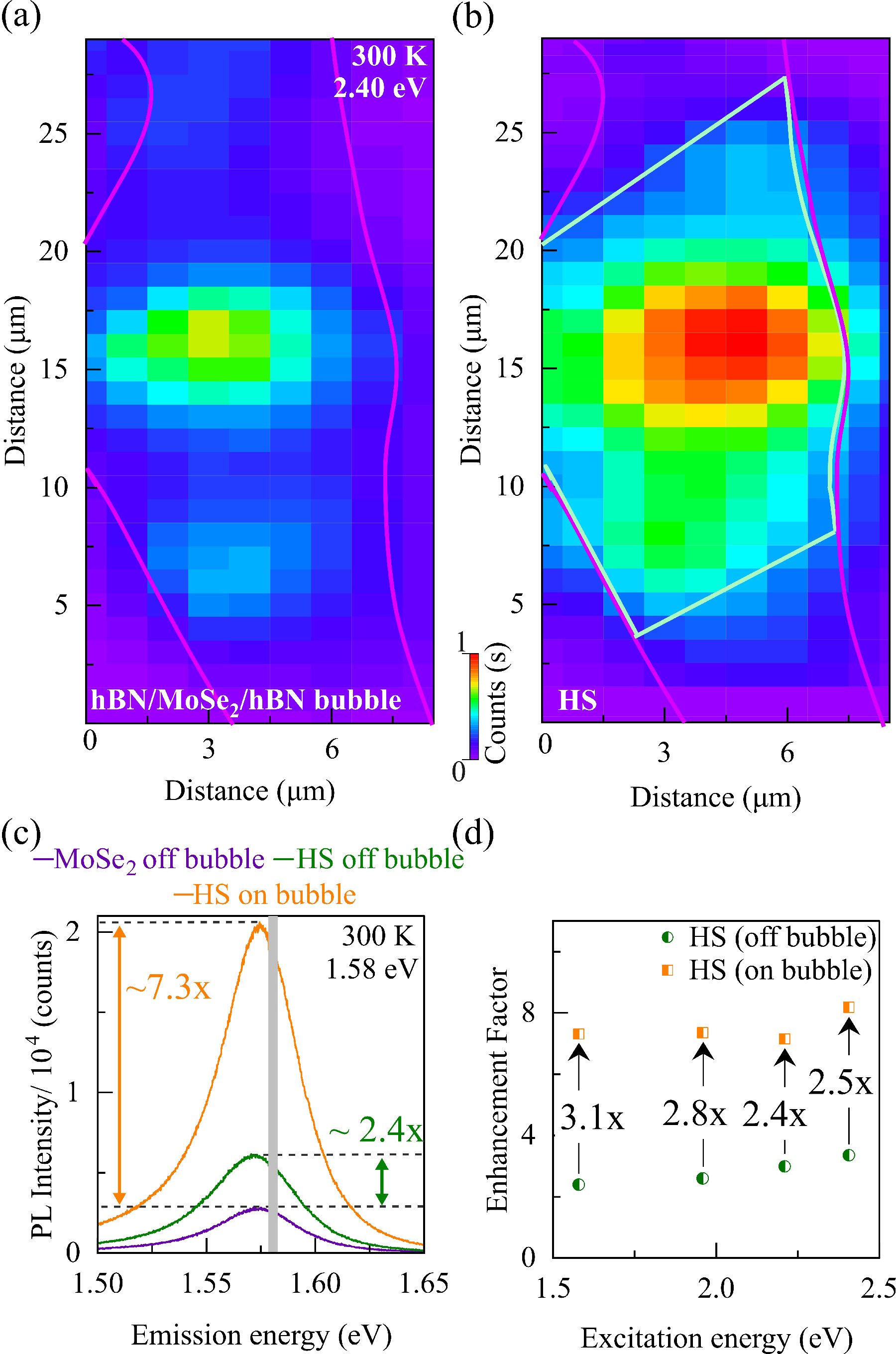}
    \caption
    {\label{Fig_2}  
    (a) Spatially resolved RT PL intensity map under 2.40 eV excitation from the white dotted area marked in ~Fig.\ref{fig_1}(c) prior to the transfer of top \ReS layer. The marked area indicates the ML \MoSe region. \MoSe emission is stronger in the bubble region than in the surrounding flat region.
    (b) PL intensity map of the same region under the same experimental conditions, after placing the top \ReS layer. HS area at the bubble apex shows the maximum PL intensity.
    (c) RT PL spectra measured under 1.58 eV excitation, from ML \MoSe, on bubble and off bubble HS areas show $\sim$$7.3\times$ and $\sim$$2.4\times$ enhancement, respectively. The gray shaded line indicates the spectral range blocked by the Bragg filter to suppress the leaking from the excitation laser.
    (d) PL enhancement factor as a function of the excitation energy. The maximum PL enhancement factor between the HS regions of $\sim$$3\times$ was observed under 1.58 eV excitation.}  
\end{figure} 

%%%%%%%%%%%%%%%%%{Photoluminescence Mapping and Energy Transfer Enhancement}

To probe the effect of this localized strain on the optical response, RT PL mapping is performed under 2.40~eV excitation. 
In order to decouple any optical interference effects arising from the hBN bubble/sapphire cavity, PL measurements are first carried out on the partial stack (hBN/\MoSe/hBN bubble/sapphire) before transferring the top \ReS layer [Fig.~\ref{Fig_2}(a)]. 
The map reveals a localized enhancement of the MoSe$_2$ emission at the bubble apex. 
This enhancement is attributed to the strain-induced funneling effect~\cite{StrainBubbles2019ACSPhotonics,Xiong2024MoS2BubblesHER}. 
The tensile strain generates a spatial variation in the \MoSe optical-transition energy, as indicated by the PL energy redshift (SI Section S4).
%%%%% MoSe2 optical-transition energy, as indicated by the spatial PL redshift.
As a result, photoexcited excitons drift toward the region of maximum tensile strain, increasing the local exciton population and consequently enhancing PL emission from \MoSe ~\cite{StrainBubbles2019ACSPhotonics,Xiong2024MoS2BubblesHER}. 
After the transfer of the top ML \ReS to complete the HS, the PL map of the same region is re-acquired [Fig.~\ref{Fig_2}(b)]. 
The PL intensity map clearly shows that the HS emission in the strained region is significantly enhanced.
Quantitative PL spectra under 1.58~eV excitation comparing the three regions are shown in Fig.~\ref{Fig_2}(c). 
This excitation energy is chosen to selectively populate the A exciton of \MoSe, reducing the inter-/intravalley carrier scattering to increase the ET efficiency. 
The \MoSe emission in the unstrained HS is enhanced by a factor of $\sim$$2.4\times$ relative to the ML \MoSe layer [Fig.~\ref{Fig_2}(c)] --- demonstrating an ET process from the \ReS to the \MoSe layer~\cite{Karmakar2022I}. 
In the strained HS, this enhancement factor increases to $\sim$$7.3\times$ [Fig.~\ref{Fig_2}(c)], demonstrating that localized tensile strain enhances the ET efficiency. 
The PL enhancement factor here is defined as the integrated \MoSe PL intensity in the HS normalized to the ML \MoSe layer.  
The reported enhancement factors are corrected for the attenuation of the excitation beam caused by the absorption of the top ML ReS$_2$ in HS (SI Section S5). 
The possible contributions of exciton funneling, interlayer ET and optical-interference effects are examined in the later sections. We would like to point out that under the present experimental conditions, no distinct \ReS PL emission is resolved, and the detected HS emission is therefore dominated by the \MoSe spectral contribution~\cite{Karmakar2022I}. 

Additional PL measurements are performed using excitation energies of 1.96, 2.21 and 2.40 eV (SI Section S5 and Fig.~\ref{Fig_2}(d)).
%   $i.e.$, above bandgap excitation, remains almost similar [Fig.~\ref{Fig_2}(d)] due to the higher carrier scattering via inter/intravalley processes, resulting in a lower number of exciton population present at the band minima for the ET process. 
The maximum HS PL enhancement factor of $\sim$3$\times$ under 1.58~eV excitation in Fig.~\ref{Fig_2}(d) is attributed to the combination of exciton funneling, reduced inter-/intravalley carrier scattering, and suppression of non-radiative losses --- resulting in maximizing the ET efficiency.
A similar enhancement of the HS emission is consistently observed in two supporting samples, confirming the reproducibility of the observed effect (SI Section S6).
After correcting for the optical absorption at each excitation energy, the relative PL QY of the strained HS is enhanced by $\sim8\times$, $7\times$, $7\times$, and $8\times$ (SI Section S7) compared to the ML \MoSe at excitation energies of 1.58, 1.96, 2.21, and 2.40 eV, respectively.

%%%%%%%%%%%%%%%%%{Band Structure calculations}

\begin{figure}[t!]
    \centering
    \includegraphics[width=1\linewidth]{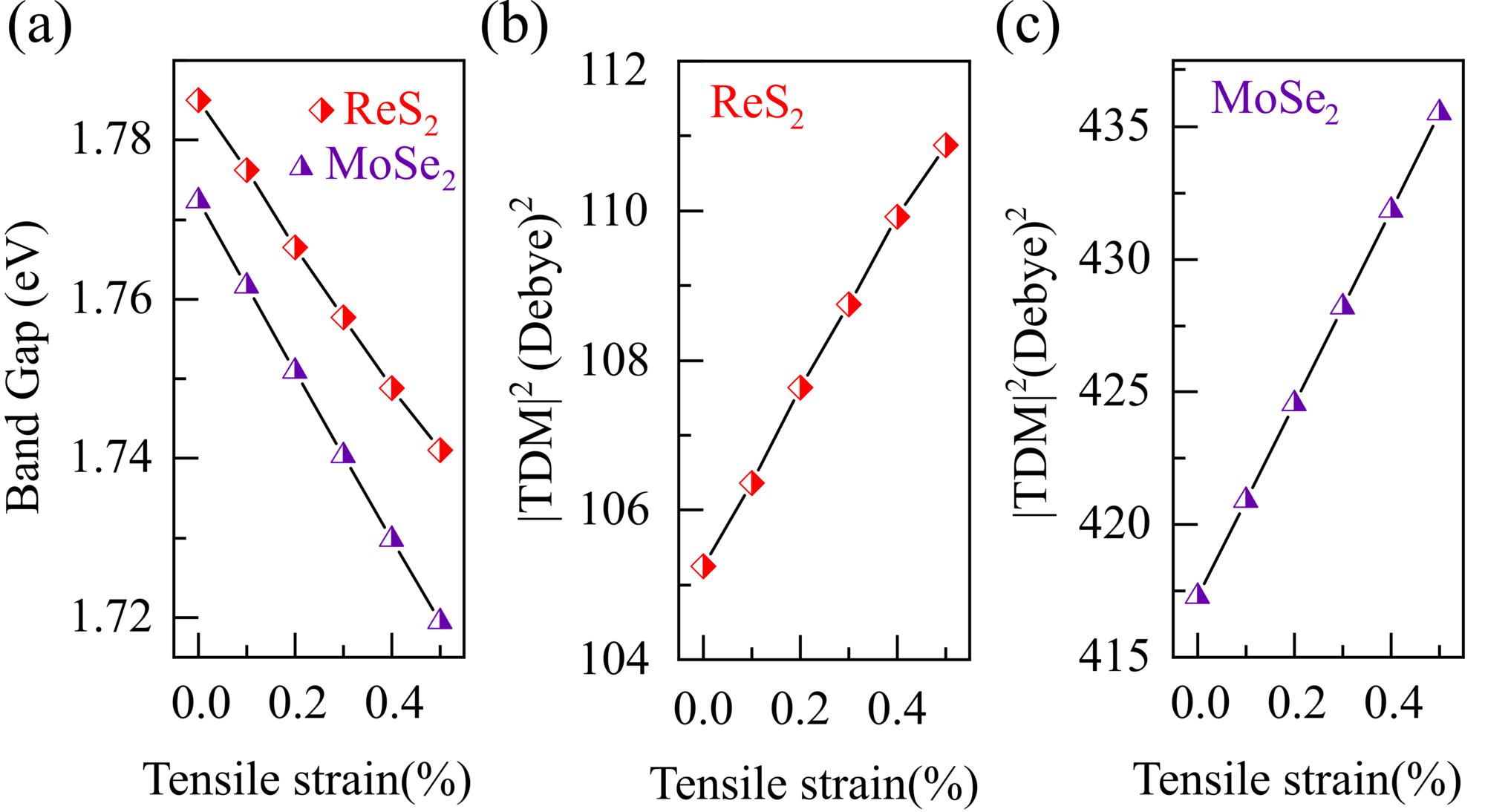}
    \caption
    {\label{fig_3}
(a) First-principles DFT  calculations of the electronic band structures of both TMDCs showing a continuous reduction of the bandgaps with an increasing biaxial tensile strain. The solid lines are guides to the eye. \ReS always maintains a higher bandgap than \MoSe throughout this range.
(b--c) Calculated $|\mathrm{TDM}|^2$ for MLs of \ReS and \MoSe, respectively, as a function of biaxial tensile strain. $|\mathrm{TDM}|^2$ shows a steady increase with the tensile strain.
    }
    \end{figure}

%%%%%%Figure4

    \begin{figure*}[t!]
    \centering
    \includegraphics[width=1\linewidth]{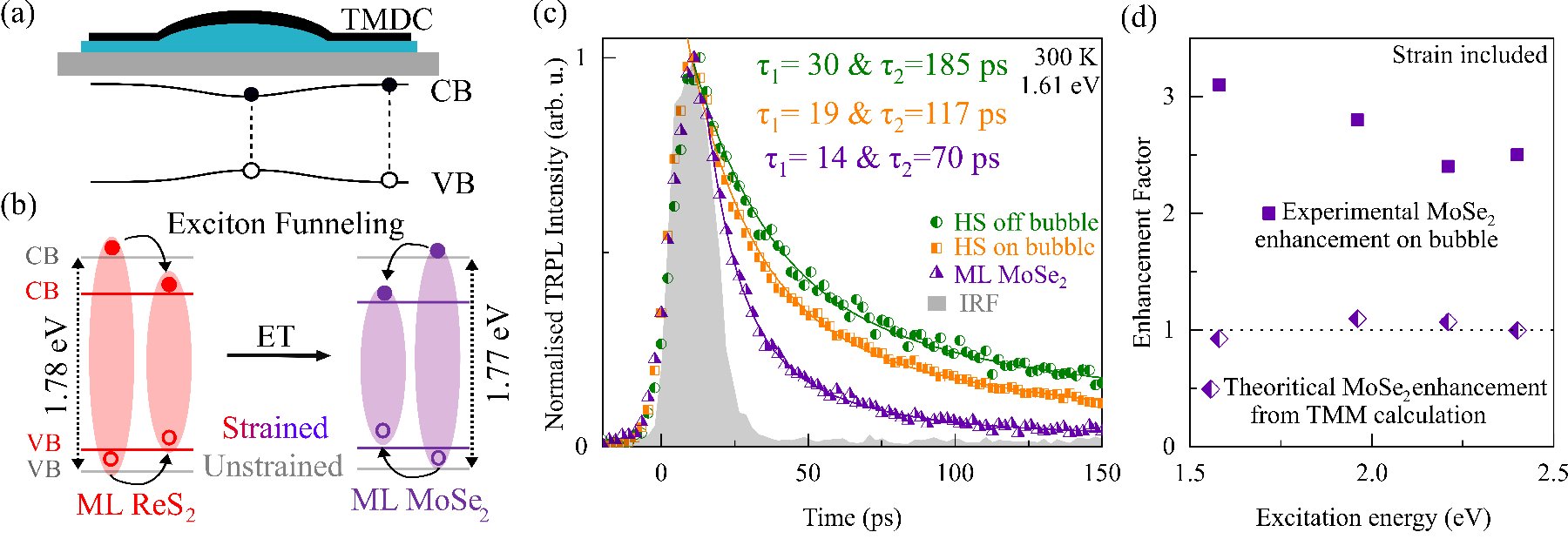}
    \caption
    {\label{fig_4}
    (a,b) Schematic illustration of strain-induced bandgap modulation and exciton funneling and the proposed dipole-mediated ET pathway from \ReS to \MoSe.  
    (c) RT TRPL spectra acquired from the ML \MoSe, and the strained and unstrained HS area under 1.61 eV excitation. The gray shaded region shows the instrument response function (IRF). The decay curves are fitted using a biexponential decay function to extract the fast ($\tau_1$) and slow ($\tau_2$) decay components. The strained HS exhibits a faster decay components than the unstrained HS.
    (d) Strain dependent TMM calculation of the \MoSe optical enhancement factor shows a near unity value for the entire excitation range. The adopted model shows that the optical interference effect from the bubble structure is insufficient to account for the measured PL enhancement.
%Strain included TMM calculations of the optical-interference induced enhancement of the \MoSe emission, compared with the experimentally measured enhancement factor. This proves that the optical interference from the bubble structure has no contribution in the reported PL enhancement.
    }
    \end{figure*} 
%%%%%%%%%%%%%%%%%{Band Structure explanations}   

To gain theoretical insight into the enhanced MoSe$_2$ emission, first-principles electronic-structure calculations are performed for biaxially strained ML ReS$_2$ and ML MoSe$_2$ (details in SI Section S8). 
The evolution of the electronic direct gaps with biaxial tensile strain is summarized in Fig.~\ref{fig_3}(a), calculated at the band minima $i.e.$, $\Gamma$ valley for ReS$_2$ and the $K$ valley for MoSe$_2$~\cite{Cheng2020, Priyanka2024}. 
For the experimentally relevant strain range, both materials exhibit an approximately linear reduction in bandgap with increasing tensile strain. 
This calculated trend is qualitatively consistent with the small PL redshift observed in the bubble region (SI Section S4), indicating that the biaxial tensile strain  reduces the bandgap~\cite{Cheng2020, Priyanka2024}. 
The optical transition probability is proportional to the intralayer $|\mathrm{TDM}|^2$~\cite{Gayatri2026FastInterlayerET}. The calculated $|\mathrm{TDM}|^2$ for the ReS$_2$ and MoSe$_2$ as a function of biaxial tensile strain are shown in Figs.~\ref{fig_3}(b) and (c), respectively (SI Section S8). 
The linear increase in $|\mathrm{TDM}|^2$ for both MLs with the tensile strain, may help in enhancing the dipole-dipole coupling~\cite{Gayatri2026FastInterlayerET}. Due to a limited computational capability, only single-particle band structure is considered and a discussion related to the change in the excitonic binding energy as a function of strain is omitted. 
%As the TDM scales inversely with the transition energy, the strain-induced reduction of the bandgap results in a increase in the transition probability for both MLs. The simultaneous increase in TDM of both the donor and acceptor layers resulted in strengthening the dipole-dipole coupling~\cite{Gayatri2026FastInterlayerET}.

%%%%%%%%%%%%%%%%%{Schematics}
A schematic illustration of the bandgap modulation and exciton funneling along the bubble is presented in Fig.~\ref{fig_4}(a), where at the apex of the bubble the TMDC bandgap shows the maximum reduction due to the maximum biaxial tensile strain~\cite{StrainBubbles2019ACSPhotonics,Xiong2024MoS2BubblesHER}. 
Figure~\ref{fig_4}(b) shows the proposed ET mechanism: gray and colored bands represent the unstrained and strained band minima, respectively. At the bubble apex, excitons accumulate \textit{via} funneling and an increased TDM helps in an effective ET coupling from the \ReS to \MoSe layer. The slightly higher \ReS excitonic energy compared to \MoSe [Fig.~\ref{fig_3}a]~\cite{Karmakar2022I}, further increases the probability of dipolar coupling outside of the light cones. A combination of all these factors resulted in the observed HS PL enhancement at the bubble apex.  
%Tensile strain reduces the bandgaps. 
%The strain gradient associated with the bubble generates a local potential landscape that may drive exciton diffusion toward lower-energy regions within the strained region~\cite{StrainBubbles2019ACSPhotonics}. 
%The strain-induced exciton-energy landscape may increase the local exciton population through funneling, while the calculated changes in the donor and acceptor TDM strengthsmay modify their dipole-dipole coupling. These effects provide a possible contribution to the enhanced \MoSe emission.
%The combined effects of the exciton funneling and increased TDM, enhance the dipole-dipole coupling from ReS$_2$ to MoSe$_2$, resulting in the experimentally observed enhancement of the MoSe$_2$ emission.
%%%%%%%%%%%%%%%%%{TRPL}

To probe the influence of strain on the exciton dynamics, RT time-resolved PL (TRPL) measurements are performed on isolated ML \MoSe, and on both the HS regions using an excitation energy of 1.61~eV [Fig.~\ref{fig_4}(c)]. 
The decay curves are fitted with a biexponential decay function, yielding a fast ($\tau_1$) and a slow ($\tau_2$) decay component. 
$\tau_1$ is associated with the intrinsic PL decay, whereas $\tau_2$ reflects the slower nonradiative recombination processes~\cite{Tanoh2020}. 
We will thus focus only on $\tau_1$ in our discussion. 
For isolated ML MoSe$_2$, $\tau_1=13.8\pm0.3$ ps. 
In the strained and unstrained HS regions, the corresponding values are $19.0\pm0.3$ and $29.9\pm0.8$ ps, respectively.
The slower $\tau_1$ in the unstrained HS as compare to ML \MoSe is a typical signature of an ET process~\cite{ Karmakar2024TwistedMoSe2,Gayatri2026FastInterlayerET}. 
In comparison, a shorter $\tau_1$ observed for the strained HS (but comparable with that of ML \MoSe); indicating a faster exciton recombination under near-resonant excitation. 
This behavior is attributed to the strain-induced bandgap narrowing, which increases the exciton funneling and enhances the radiative recombination \textit{via} ET process. 
In contrast, under above-bandgap excitation at 2.27~eV (SI Section S9), the decay dynamics of the strained and unstrained HS become nearly identical. 
At these higher excitation energies, carrier relaxation through phonon-assisted inter-/intravalley scatterings dominate the exciton dynamics, largely masking the comparatively small influence of strain-induced exciton funneling. 
It is important to mention that, since the HS PL emission is dominated by the \MoSe signature, we assigned the HS TRPL spectra only for the \MoSe contribution. 
Resolving the \ReS emission from the TRPL data is beyond our instrumental resolution, a separate ultrafast study is required for that.
%%%%%%%%%%%%%%%%%{TMM}

Finally, to evaluate whether the enhanced \MoSe emission could partly originate from the bubble-induced optical interference, the optical response of the HS is modeled using a coherent TMM method~\cite{Troparevsky2010TransferMatrix, Mitsas1995GeneralizedMatrix, Byrnes2016MultilayerOptical}. The model consists of an air/ML ReS$_2$/hBN/ML MoSe$_2$/hBN bubble/sapphire stack (details are provided in SI Sections S10-S12). 
Our model accounts for multiple reflections, optical interference effects, and absorption within the multilayer stack. 
The theoretically calculated \MoSe optical enhancement factor shown in Fig.~\ref{fig_4}(d) remains close to unity over the investigated excitation-energy range, indicating that the bubble geometry produces only a weak modulation of the optical absorption.
A similar calculation for the ReS$_2$ layer, presented in Section S13 of the SI, leads to the same conclusion that optical interference is substantially smaller than the measured PL enhancement.
Within the adopted model, the optical interference alone is therefore insufficient to account for the pronounced HS PL enhancement. 
The remaining enhancement is a result of an enhanced ET process due to a combination of strain-modified exciton dynamics and exciton funneling. However, the individual contributions of these processes cannot be quantitatively separated by the present measurements.
%%%%%%%%%%%%%%%%%% {Conclusions}
\section{Conclusions \label{sec:Conclusions}}
In summary, we demonstrate that only $\sim$0.1\% biaxial tensile strain produces an $\sim$8-fold increase in the relative MoSe$_2$ PL QY in MLs ReS$_2$/hBN/MoSe$_2$ HS. The thin ($\sim$12 nm) hBN spacer suppresses the CT process.
Raman and PL measurements consistently indicate weak tensile strain, while first-principles calculations reveal strain-induced bandgap narrowing and increased TDMs that strengthen dipole–dipole coupling. 
The strain gradient further promotes the exciton funnelling toward the bubble apex. TMM calculation rules out optical interference as the dominant origin of the enhancement. Importantly, these results are obtained in a purely ML HS; increasing the optical absorption of the donor layer is therefore expected to enable even larger gains. Our findings establish small tensile strain as a practical and controllable route to engineer ET and PL QY in ultrathin vdW optoelectronic devices.

%{Methods}
\section{Methods\label{sec:Methods}}

\subsection{Sample Preparation}
\textit{hBN Growth on Sapphire Substrate:} The hBN layers were grown on 2-inch c-plane sapphire substrates with a 0.3$^\circ$ offcut using metal-organic vapor phase epitaxy (MOVPE) in an Aixtron CCS 3 $\times$ 2 system equipped with an ARGUS thermal mapping system. Triethylboron and ammonia were used as boron and nitrogen precursors, respectively, while hydrogen served as the carrier gas. The growth was performed at 1400 $^\circ$C using a two-stage epitaxial growth process described in Ref.~\cite{Dabrowska2020hBNMOVPE}.

\textit{hBN Bubble Formation:} Localized bubbles were generated by electron-beam exposure using an FEI Helios 600 Dual Beam system equipped with a Raith Elphy electron lithography setup. During the exposure, the electron beam current and acceleration voltage were maintained at 1~nA and 5~keV, respectively.
The resulting bubbles are mechanically stable under ambient conditions and retain their geometry over extended periods.  Previous study on analogous hBN bubbles reported tensile strains of approximately $0.1\%$~\cite{Tatarczak2025}, consistent with the strain determined in the present work from Raman and PL measurements.

\textit{HS Fabrication:} hBN crystals were obtained from the National Institute for Materials Science (NIMS), Japan, while bulk MoSe$_2$ and ReS$_2$ crystals were purchased from 2D Semiconductors, USA. ML TMDCs and thin hBN flakes were mechanically exfoliated onto polydimethylsiloxane (PDMS) gel films. The layer thicknesses were confirmed by the optical microscope, PL and Raman spectroscopy measurements.
The vertical stacking of the layers was carried out using a deterministic dry-transfer technique. After each layer is transferred, the samples were annealed inside an e-beam evaporator chamber under high vacuum ($\sim$$7 \times 10^{-4}$ Torr) at 180 $^\circ$C for 3~h under argon flow of 50~sccm to improve the interlayer adhesion and interface quality.

\subsection{Characterization}
\textit{Optical Characterization:} PL measurements were performed using Cobolt diode lasers with excitation energies of 1.58, 1.96, 2.21, and 2.40~eV. Raman scattering measurements were carried out using the same setup under 1.96~eV excitation. The laser was focused onto the sample using a Mitutoyo Plan Apo SL 50$\times$ objective lens (NA~= 0.55), producing a spot diameter of $\sim$1~$\mu$m. The emitted signal was collected in a backscattering geometry through the same objective and directed to a Teledyne Princeton Instruments spectrometer equipped with a 150~grooves/mm grating and a CCD detector. The average excitation power was maintained at $\sim$30~$\mu$W.

\textit{AFM Scan:} AFM measurements were performed using a Dimension Icon microscope equipped with a NanoScope 6 controller (Bruker Corporation, Billerica, MA, USA). Topographical images were acquired in PeakForce Tapping mode using ScanAsyst-Air probes. The probes were calibrated using the thermal tuning method prior to measurements.

\textit{TRPL Measurements:} TRPL investigations were performed using an S20 synchroscan streak camera system. The samples were excited under 1.61 and 2.27~eV excitation using the MIRA laser system and the second harmonic of a Ti:sapphire fs oscillator. The average excitation power was kept at $\sim$26~$\mu$W with a spot diameter of $\sim$1~$\mu$m.

\section{Author Contributions}
A.K. conceived the project and provided the necessary funding. 
G., M.B., P.T., A.K.D., J.B., and A.W. prepared the hBN bubbles. 
G. and A.K. fabricated the samples. 
G. and M.R.M. performed the optical experiments. 
G., M.R., T.K., and P.K. carried out the TRPL experiments. 
G. carried out the data analysis with inputs taken from A.K. and M.R.M. 
M.A., D.D., and S.K.N. conducted the theoretical studies.
G., A.K., M.R.M., and A.B. interpreted the experimental results. 
T.T. and K.W. provided the hBN crystals. 
G., A.K., and M.R.M. wrote the manuscript. 
All authors were consulted and their feedback was incorporated before submission.

\section{Acknowledgments}
This work is supported by the National Science Centre, Poland (Grant No. 2022/47/D/ST3/02086).
D.D. and S.K.N. would like to appreciate the funding by the DST project (RP-312), and also acknowledge the access of the high-performance computing facility provided by the Institute of Physics, Bhubaneswar and Indian Institute of Technology Bhubaneswar.
P.T., A.W. and J.B. thank NCN for the support (Grant No. 2022/45/N/ST7/03355).
K.W. and T.T. acknowledge support from the JSPS KAKENHI (Grant Numbers 21H05233 and 23H02052), the CREST (JPMJCR24A5), JST and World Premier International Research Center Initiative (WPI), MEXT, Japan.  ChatGPT (GPT 5.5) and Grok 4.5 have been used for language editing and paraphrasing.

\section {Competing interests}
We declare no competing financial interests.

\section{Supporting Information}
Additional experimental details; AFM characterization, strain analysis from PL measurements; PL measurements under different excitation energies; relative PL quantum yield calculations; DFT calculations; TRPL measurements and analysis; and transfer-matrix calculations. The Supporting Information is available online.

\section{Data availability}
The data that support the findings of this study are available from the corresponding author upon reasonable request.

\bibliographystyle{achemso}
\bibliography{biblio}

@article{Manzeli2017,
  author = {Manzeli, S. and Ovchinnikov, D. and Pasquier, D. and Yazyev, O. V. and Kis, A.},
  title = {2D Transition Metal Dichalcogenides},
  journal = {Nat. Rev. Mater.},
  volume = {2},
  pages = {17033},
  year = {2017},
  doi = {10.1038/natrevmats.2017.33}
}

@article{Qi2023StrainEngineering,
  author    = {Qi, Zhaowei and Zheng, Meng and Wu, Zhiyuan and Jiang, Yong and Chen, Yong P. and Chen, Yong},
  title     = {Recent Progress in Strain Engineering on van der Waals 2D Materials: Tunable Electrical, Optical, Magnetic, and Topological Properties},
  journal   = {Adv. Mater.},
  year      = {2023},
  volume    = {35},
  number    = {6},
  pages     = {2205714},
  doi       = {10.1002/adma.202205714},
  url       = {https://doi.org/10.1002/adma.202205714},
  publisher = {Wiley}
}

@article{Conley2013,
  author = {Conley, H. J. and Wang, B. and Ziegler, J. I. and Haglund, R. F. and Pantelides, S. T. and Bolotin, K. I.},
  title = {Bandgap Engineering of Strained Monolayer and Bilayer MoS$_2$},
  journal = {Nano Lett.},
  volume = {13},
  pages = {3626--3630},
  year = {2013},
  doi = {10.1021/nl4014748}
}

@article{Ai2025,
  author = {Ai, Ruoqi and Cui, Ximin and Li, Yang and Zhuo, Xiaolu},
  title = {Local Strain Engineering of Two-Dimensional Transition Metal Dichalcogenides Towards Quantum Emitters},
  journal = {Nano-Micro Lett.},
  volume = {17},
  pages = {104},
  year = {2025},
  doi = {10.1007/s40820-024-01611-1}
}

@article{Branny2017,
  author = {Branny, Artur and Kumar, Santosh and Proux, Rapha{\"e}l and Gerardot, Brian D.},
  title = {Deterministic Strain-Induced Arrays of Quantum Emitters in a Two-Dimensional Semiconductor},
  journal = {Nat. Commun.},
  volume = {8},
  pages = {15053},
  year = {2017},
  doi = {10.1038/ncomms15053}
}

@article{Datta2020,
  author = {Datta, Souvik and Strachan, Douglas R.},
  title = {Strain-Induced Exciton Localization in Two-Dimensional Semiconductors},
  journal = {ACS Nano},
  volume = {14},
  number = {10},
  pages = {13658--13668},
  year = {2020},
  doi = {10.1021/acsnano.0c05730}
}

@article{Kumar2015,
  author = {Kumar, Santosh and Kaczmarczyk, Artur and Gerardot, Brian D.},
  title = {Strain-Induced Spatial and Spectral Isolation of Quantum Emitters in Mono- and Bilayer WSe$_2$},
  journal = {Nano Lett.},
  volume = {15},
  number = {11},
  pages = {7567--7573},
  year = {2015},
  doi = {10.1021/acs.nanolett.5b03312}
}

@article{StrainBubbles2019ACSPhotonics,
  author = {Tyurnina, Anastasia V. and Bandurin, Denis A. and Khestanova, Ekaterina and Kravets, Vasyl G. and Koperski, Maciej and Guinea, Francisco and Grigorenko, Alexander N. and Geim, Andre K. and Grigorieva, Irina V.},
  title = {Strained Bubbles in van der Waals Heterostructures as Local Emitters of Photoluminescence with Adjustable Wavelength},
  journal = {ACS Photonics},
  volume = {6},
  number = {2},
  pages = {516--524},
  year = {2019},
  doi = {10.1021/acsphotonics.8b01497},
  publisher = {American Chemical Society}
}

@article{Xiong2024MoS2BubblesHER,
  author  = {Xiong, Jie and Xiong, Junjie and Cai, Yuchen and Dong, Wenlong and Luo, Xinying and Yu, Zhongliang and Liu, Bowen and Liu, Luqi and Liang, T. and Huang, Demei and Wang, Zhenxing and Gao, Yang and Wang, Bin},
  title   = {Strain derived from bubbles at monolayer MoS2/hBN interfaces for enhanced hydrogen evolution reaction activity},
  journal = {Chem Catalysis},
  year    = {2024},
  volume  = {4},
  number  = {4},
  pages   = {100951},
  doi     = {10.1016/j.checat.2024.100951},
  publisher = {Elsevier}
}

@article{Feng2012,
  author = {Feng, Ji and Qian, Xiaofeng and Huang, Cheng-Wei and Li, Ju},
  title = {Strain-Engineered Artificial Atom as a Broad-Spectrum Solar Energy Funnel},
  journal = {Nat. Photonics},
  volume = {6},
  pages = {866--872},
  year = {2012},
  doi = {10.1038/nphoton.2012.285}
}

@Article{Mueller2018,
  author = {Mueller, Thomas and Malic, Ermin},
  title = {Exciton physics and device application of two-dimensional transition metal dichalcogenide semiconductors},
  journal = {npj 2D Mater. Appl.},
  year = {2018},
  volume = {2},
  number = {1},
  pages = {29},
  doi = {10.1038/s41699-018-0074-2},
  url = {https://doi.org/10.1038/s41699-018-0074-2}
}

@article{Geim2013,
  author = {Geim, A. K. and Grigorieva, I. V.},
  title = {Van der Waals Heterostructures},
  journal = {Nature},
  volume = {499},
  pages = {419--425},
  year = {2013}
}

@article{Rivera2015,
  author = {Rivera, P. and Schaibley, J. R. and Jones, A. M. and Ross, J. S. and Wu, S. and Aivazian, G. and Klement, P. and Seyler, K. and Yan, J. and Mandrus, D. G. and Yao, W. and Xu, X.},
  title = {Observation of Long-Lived Interlayer Excitons in Monolayer MoSe2/WSe2 Heterostructures},
  journal = {Nat. Commun.},
  volume = {6},
  pages = {6242},
  year = {2015}
}

@article{Novoselov2016,
  author = {Novoselov, Kostya S. and Mishchenko, Artem and Carvalho, A. and Neto, A. H. Castro},
  title = {2D Materials and van der Waals Heterostructures},
  journal = {Science},
  volume = {353},
  number = {6298},
  pages = {aac9439},
  year = {2016},
  doi = {10.1126/science.aac9439}
}

@article{Jin2018UltrafastDynamics,
  author = {Jin, Chenhao and Ma, Eric Yue and Karni, Ouri and Regan, Emma C. and Wang, Feng and Heinz, Tony F.},
  title = {Ultrafast Dynamics in van der Waals Heterostructures},
  journal = {Nat. Nanotechnol.},
  volume = {13},
  number = {11},
  pages = {994--1003},
  year = {2018},
  doi = {10.1038/s41565-018-0298-5}
}

@article{Kozawa2016FastET,
  author = {Kozawa, Daichi and Carvalho, Alexandra and Verzhbitskiy, Ivan and Giustiniano, Francesco and Miyauchi, Yuhei and Mouri, Shinichiro and Castro Neto, A. H. and Matsuda, Kazunari and Eda, Goki},
  title = {Evidence for Fast Interlayer Energy Transfer in MoSe$_2$/WS$_2$ Heterostructures},
  journal = {Nano Lett.},
  volume = {16},
  number = {7},
  pages = {4087--4093},
  year = {2016},
  doi = {10.1021/acs.nanolett.6b00801}
}

@article{Karmakar2024TwistedMoSe2,
  author = {Karmakar, Arka and Al-Mahboob, Abdullah and Zawadzka, Natalia and Raczy{\'n}ski, Mateusz and Yang, Weiguang and Arfaoui, Mehdi and Gayatri and Kucharek, Julia and Sadowski, Jerzy T. and Shin, Hyeon Suk and Babi{\'n}ski, Adam and Pacuski, Wojciech and Kazimierczuk, Tomasz and Molas, Maciej R.},
  title = {Twisted MoSe$_2$ Homobilayer Behaving as a Heterobilayer},
  journal = {Nano Lett.},
  volume = {24},
  number = {31},
  pages = {9459--9467},
  year = {2024},
  doi = {10.1021/acs.nanolett.4c01764}
}

@article{Liu2021EnergyChargeTransfer,
  author = {Liu, Junyi and Li, Zi and Zhang, Xu and Lu, Gang},
  title = {Unraveling Energy and Charge Transfer in Type-II van der Waals Heterostructures},
  journal = {npj Comput. Mater.},
  volume = {7},
  pages = {191},
  year = {2021},
  doi = {10.1038/s41524-021-00663-w}
}

@article{Aftab2023EnergyChargeTransfer,
  author = {Aftab, Sikandar and Iqbal, Muhammad Zahir and Hegazy, Hosameldin Helmy and Azam, Sikander and Kabir, Fahmid},
  title = {Trends in Energy and Charge Transfer in 2D and Integrated Perovskite Heterostructures},
  journal = {Nanoscale},
  volume = {15},
  number = {8},
  pages = {3610--3629},
  year = {2023},
  doi = {10.1039/D2NR07141J}
}

@article{Gayatri2026FastInterlayerET,
  author = {Gayatri and Arfaoui, Mehdi and Das, Debashish and Kazimierczuk, Tomasz and Ayari, Sabrine and Zawadzka, Natalia and Taniguchi, Takashi and Watanabe, Kenji and Babi{\'n}ski, Adam and Nayak, Saroj K. and Molas, Maciej R. and Karmakar, Arka},
  title = {Fast Interlayer Energy Transfer from the Lower Bandgap MoS$_2$ to the Higher Bandgap WS$_2$},
  journal = {npj 2D Mater. Appl.},
  volume = {10},
  pages = {25},
  year = {2026},
  doi = {10.1038/s41699-026-00661-w}
}

@article{Kim2022StrainCTET,
  author = {Joon-Seok Kim and Nikhilesh Maity and Myungsoo Kim and Suyu Fu and Rinkle Juneja and Abhishek Singh and Deji Akinwande and Jung-Fu Lin},
  title = {Strain-Modulated Interlayer Charge and Energy Transfers in MoS$_2$/WS$_2$ Heterobilayer},
  journal = {ACS Appl. Mater. Interfaces},
  year = {2022},
  volume = {14},
  number = {41},
  pages = {46841--46849},
  doi = {10.1021/acsami.2c10982},
  url = {https://doi.org/10.1021/acsami.2c10982},
  publisher = {American Chemical Society}
}

@article{Cheng2020,
  author = {Cheng, Xuerui and Jiang, Liying and Li, Yuanyuan and Zhang, Huanjun and Hu, Chuansheng and Xie, Shiyu and Liu, Miao and Qi, Zeming},
  title = {Using Strain to Alter the Energy Bands of the Monolayer MoSe$_2$: A Systematic Study Covering Both Tensile and Compressive States},
  journal = {Appl. Surf. Sci.},
  volume = {521},
  pages = {146398},
  year = {2020},
  doi = {10.1016/j.apsusc.2020.146398}
}

@article{Priyanka2024,
  author = {Priyanka and Ritu and Vinod Kumar and Ramesh Kumar and Fakir Chand},
  title = {Tailoring the Electronic and Optical Properties of ReS$_2$ Monolayer Using Strain Engineering},
  journal = {Micro Nano Struct.},
  volume = {192},
  pages = {207873},
  year = {2024},
  doi = {10.1016/j.micrna.2024.207873}
}

@article{Binder2023Epitaxial,
  author = {Binder, Johannes and
            D{\k{a}}browska, Aleksandra Krystyna and
            Tokarczyk, Mateusz and
            Ludwiczak, Katarzyna and
            Bo{\.z}ek, Rafa{\l} and
            Kowalski, Grzegorz and
            St{\k{e}}pniewski, Roman and
            Wysmo{\l}ek, Andrzej},
  title = {Epitaxial Hexagonal Boron Nitride for Hydrogen Generation by Radiolysis of Interfacial Water},
  journal = {Nano Lett.},
  year = {2023},
  volume = {23},
  number = {4},
  pages = {1267--1272},
  doi = {10.1021/acs.nanolett.2c04434},
  publisher = {American Chemical Society}
}

@article{Tatarczak2025,
  author = {Tatarczak, Piotr and F{\k{a}}s, Tomasz and Paw{\l}owski, Jan and D{\k{a}}browska, Aleksandra Krystyna and Suffczy{\'n}ski, Jan and Tokarczyk, Mateusz and Wr{\'o}bel, Piotr and Wysmo{\l}ek, Andrzej and Binder, Johannes},
  title = {Deterministic hBN Bubbles as a Versatile Platform for Studies on Single-Photon Emitters},
  journal = {Adv. Funct. Mater.},
  year = {2025},
  volume = {36},
  number = {24},
  pages = {e26312},
  doi = {10.1002/adfm.202526312},
  publisher = {Wiley-VCH}
}

@article{Khestanova2016UniversalShape,
  author = {Khestanova, Ekaterina and Guinea, Francisco and Fumagalli, Luca and Geim, Andre K. and Grigorieva, Irina V.},
  title = {Universal Shape and Pressure Inside Bubbles Appearing in van der Waals Heterostructures},
  journal = {Nat. Commun.},
  volume = {7},
  pages = {12587},
  year = {2016},
  doi = {10.1038/ncomms12587}
}

@article{Hwang2022ShearStrainPL,
  author = {Hwang, Hyeong-Yong and Kim, Jihye and Lee, Seung Hwan and Kim, Hyeonjin and Kim, Seongjoon and Lee, Jieun},
  title = {Shear-Strain-Mediated Photoluminescence Manipulation in Two-Dimensional Transition Metal Dichalcogenides},
  journal = {2D Mater.},
  volume = {9},
  number = {1},
  pages = {015011},
  year = {2022},
  doi = {10.1088/2053-1583/ac351d}
}

@article{Blundo2020Engineered,
  author = {Blundo, Elena and Di Giorgio, Cinzia and Pettinari, Giorgio and Yildirim, Tanju and Felici, Marco and Lu, Yuerui and Bobba, Fabrizio and Polimeni, Antonio},
  title = {Engineered Creation of Periodic Giant, Nonuniform Strains in MoS2 Monolayers},
  journal = {Adv. Mater. Interfaces},
  year = {2020},
  volume = {7},
  number = {17},
  pages = {2000621},
  doi = {10.1002/admi.202000621}
}

@article{Karmakar2022I,
  author = {Karmakar, Arka and Al-Mahboob, Abdullah and Petoukhoff, Christopher E. and Kravchyna, Oksana and Chan, Nicholas S. and Taniguchi, Takashi and Watanabe, Kenji and Dani, Keshav M.},
  title = {Dominating Interlayer Resonant Energy Transfer in Type-II 2D Heterostructure},
  journal = {ACS Nano},
  year = {2022},
  volume = {16},
  number = {3},
  pages = {3861--3869},
  doi = {10.1021/acsnano.1c08798},
  publisher = {American Chemical Society}
}

@article{Dandu2020,
  author = {Dandu, Medha and Watanabe, Kenji and Taniguchi, Takashi and Sood, Ajay K. and Majumdar, Kausik},
  title = {Spectrally Tunable, Large Raman Enhancement from Nonradiative Energy Transfer in the van der Waals Heterostructure},
  journal = {ACS Photonics},
  volume = {7},
  number = {2},
  pages = {519--527},
  year = {2020},
  doi = {10.1021/acsphotonics.9b01648}
}

@article{Yagmurcukardes2018,
author = {Yagmurcukardes, M. and Bacaksiz, C. and Unsal, E. and Akbali, B. and Senger, R. T. and Sahin, H.},
title = {Strain mapping in single-layer two-dimensional crystals via Raman activity},
journal = {Phys. Rev. B},
volume = {97},
number = {11},
pages = {115427},
year = {2018},
doi = {10.1103/PhysRevB.97.115427}
}

@article{Chang2013,
author = {Chang, Chung-Huai and Fan, Xiaofeng and Lin, Shi-Hsin and Kuo, Jer-Lai},
title = {Orbital Analysis of Electronic Structure and Phonon Dispersion in MoS$_2$, MoSe$_2$, WS$_2$, and WSe$_2$ Monolayers under Strain},
journal = {Phys. Rev. B},
volume = {88},
pages = {195420},
year = {2013},
doi = {10.1103/PhysRevB.88.195420}
}

@article{Pak2017,
author = {Pak, Sangyeon and Lee, Juwon and Lee, Young-Woo and Jang, A.-R. and Ahn, S. and Ma, K. Y. and Cho, Y. and Hong, J. and Lee, S. and Jeong, H. Y. and Im, H. and Shin, H. S. and Morris, S. M. and Cha, S. and Sohn, J. I. and Kim, J. M.},
title = {Strain-Mediated Interlayer Coupling Effects on the Excitonic Behaviors in an Epitaxially Grown MoS$_2$/WS$_2$ van der Waals Heterobilayer},
journal = {Nano Lett.},
year = {2017},
volume = {17},
number = {9},
pages = {5634--5640},
doi = {10.1021/acs.nanolett.7b02513}
}

@article{Qian2025,
author = {Qian, Wenqi and Liu, Haiyi and Tao, Guangyi and Liu, Fangxun and Lin, Sihan and Gao, Tengteng and Wang, Xueying and Hu, Qihong and Zhang, Dalin and Xiang, Dong and Lin, Lie and Qi, Pengfei and Fang, Zheyu and Liu, Weiwei},
title = {Exciton funneling in 2D artificial potential landscapes decorated by reassembled micro-bubbles},
journal = {Mater. Futures},
volume = {4},
number = {2},
pages = {025301},
year = {2025},
doi = {10.1088/2752-5724/adc8c1}
}

@article{Troparevsky2010TransferMatrix,
author = {Troparevsky, M. Claudia and Sabirianov, Renat F. and Jaswal, S. S. and Khanna, S. N. and Tilley, R. J. D. and Zhang, S.},
title = {Transfer-Matrix Formalism for the Calculation of Optical Response in Multilayer Systems: From Coherent to Incoherent Interference},
journal = {Opt. Express},
volume = {18},
number = {24},
pages = {24715--24721},
year = {2010},
doi = {10.1364/OE.18.024715}
}

@article{Mitsas1995GeneralizedMatrix,
author = {Mitsas, C. L. and Siapkas, D. I.},
title = {Generalized Matrix Method for Analysis of Coherent and Incoherent Reflectance and Transmittance of Multilayer Structures with Rough Surfaces, Interfaces, and Finite Substrates},
journal = {Appl. Opt.},
volume = {34},
number = {10},
pages = {1678--1683},
year = {1995},
doi = {10.1364/AO.34.001678}
}

@article{Byrnes2016MultilayerOptical,
author = {Byrnes, Steven J.},
title = {Multilayer Optical Calculations},
journal = { rXiv preprint arXiv:1603.02720},
year = {2016},
doi = {10.48550/arXiv.1603.02720}
}

@article{Tanoh2020,
author = {Tanoh, Arelo O. A. and Gauriot, Nicolas and Delport, G{'e}raud and Xiao, James and Pandya, Raj and Sung, Jooyoung and Allardice, Jesse and Li, Zhaojun and Williams, Cyan A. and Baldwin, Alan and Stranks, Samuel D. and Rao, Akshay},
title = {Directed Energy Transfer from Monolayer WS$_2$ to Near-Infrared Emitting PbS--CdS Quantum Dots},
journal = {ACS Nano},
volume = {14},
number = {11},
pages = {15374--15384},
year = {2020},
doi = {10.1021/acsnano.0c05818}
}

@article{Dabrowska2020hBNMOVPE,
author = {D{\k{a}}browska, Aleksandra K. and Tokarczyk, Mateusz and Kowalski, Grzegorz and Binder, Johannes and Bo{\v{z}}ek, Rafa{\l} and Borysiuk, Jakub and St{\k{e}}pniewski, Roman and Wysmo{\l}ek, Andrzej},
title = {Two Stage Epitaxial Growth of Wafer-Size Multilayer h-BN by Metal-Organic Vapor Phase Epitaxy -- A Homoepitaxial Approach},
journal = {2D Mater.},
volume = {8},
number = {1},
pages = {015017},
year = {2020},
doi = {10.1088/2053-1583/abbd1f}
}

@article{Karmakar2023ExcitationDependent,
author = {Karmakar, Arka and Arfaoui, Mehdi and Zawadzka, Natalia and Kazi, Zakir and Babi{\'n}ski, Adam and Molas, Maciej R.},
title = {Excitation-Dependent High-Lying Excitonic Exchange via Interlayer Energy Transfer from Lower- to Higher-Bandgap 2D Material},
journal = {Nano Lett.},
volume = {23},
number = {13},
pages = {5617--5624},
year = {2023},
doi = {10.1021/acs.nanolett.3c01066}
}
\end{document}